\documentstyle{article}
\begin{document}
\newcommand{\C}{{\bf C} \!\!\!\! I}
\newcommand{\HH}{H \!\!\! I}

\newcommand{\N}{N \!\!\! I}
\newcommand{\UU}{\underline{\sqcup}}
\newcommand{\UUU}{\sqcup}
\newcommand{\Z}{Z \!\!\! Z}
\newcommand{\la}{\langle}
\newcommand{\ra}{\rangle}
\newcommand{\li}
{\begin{picture}(10,10)(0,0)
\put(0,-3){\line(0,1){10}}
\end{picture}}
\newcommand{\sq}{\begin{picture}(10,10)(0,0)
\put(0,-3){\line(0,1){10}}
\put(-2,2){${}_{\Box}$}
\end{picture}}
\newcommand{\bl}
{\begin{picture}(10,10)(0,0)
\put(0,-3){\line(0,1){10}}
\put(-2,0){$\bullet$}
\end{picture}}
%
\newcommand{\dbl}
{\begin{picture}(10,15)(0,0)
\put(0,-3){\line(0,1){15}}
\put(-2,0){$\bullet$}
\put(-2,4){$\bullet$}
\end{picture}}
\newcommand{\dsq}
{\begin{picture}(10,15)(0,0)
\put(0,-3){\line(0,1){15}}
\put(-2,1){${}_\Box$}
\put(-2,7){${}_\Box$}
\end{picture}}
\newcommand{\cus}{\bigcup_{\!\!\!\! \Box}}
\newcommand{\cub}{\bigcup_{\!\!\!\! \bullet}}
\newcommand{\sms}{
\smile \!\! \circ}
\newcommand{\smb}{\smile \!\!
\bullet}
\newcommand{\rec}{\begin{picture}(0,0)(0,0)
\put(-15,60){\dashbox{1.0}(140,40){}}
\end{picture}}
\newcommand{\Cup}{
\begin{picture}(20,15)(-7,-7)
\put(0,0){\oval(10,15)[b]}
\end{picture}}
\newcommand{\Cap}{
\begin{picture}(20,15)(-7,-7)
\put(0,-10){\oval(10,15)[t]}
\end{picture}}
\newcommand{\Cus}{
\begin{picture}(20,20)(-7,-7)
\put(0,0){\oval(10,15)[b]}
\put(-3,-8){${}_{\Box}$}
\end{picture}}
\newcommand{\Cub}{
\begin{picture}(20,20)(-7,-7)
\put(0,0){\oval(10,15)[b]}
\put(-3,-10){$\bullet$}
\end{picture}}
\newcommand{\scus}{
\begin{picture}(20,20)(-7,-7)
\put(0,0){\oval(10,10)[b]}
\put(0,0){\oval(30,20)[b]}
\put(0,-10){${}_{\Box}$}
\end{picture}}
\newcommand{\scup}{
\begin{picture}(40,15)(-15,-7)
\put(0,0){\oval(10,10)[b]}
\put(0,0){\oval(30,20)[b]}
\end{picture}}
\newcommand{\scub}{
\begin{picture}(20,20)(-7,-7)
\put(0,0){\oval(10,10)[b]}
\put(0,0){\oval(30,20)[b]}
\put(0,-12){$\bullet$}
\end{picture}}
\newtheorem{th}{Theorem}
\newtheorem{cor}{Corollary}[th]
\newtheorem{de}{Definition}
\newtheorem{pr}{Proposition}
\newtheorem{co}{Corollary}[pr]
\newtheorem{rem}{Remark}

USC-93/010 February 12, 1993
\bigskip
\begin{center}
{\bf ALGEBRAS IN HIGHER DIMENSIONAL STATISTICAL MECHANICS
\\
- THE EXCEPTIONAL PARTITION (MEAN FIELD) ALGEBRAS}
\\
 Paul  Martin
\footnote{
Math. Dept., City University,  London EC1V 0HB, UK.}
\renewcommand{\thefootnote}{\dag}
 and Hubert Saleur \footnote{
Phys. Dept. and Math. Dept., University of Southern California, Los Angeles CA
 90089, USA. Work supported by the Packard Foundation.}
\end{center}

\vspace{.1in}

\begin{abstract}
We determine the structure of the partition algebra $P_n(Q)$
(a generalized Temperley-Lieb algebra) for specific
values of $Q \in \C$, focusing on the quotient which gives rise
to the partition function of $n$ site $Q$-state Potts models (in the
continuous $Q$ formulation) in arbitrarily high lattice dimensions
(the mean field case).
The algebra is non-semi-simple iff $Q$ is a non-negative integer less than $n$.
We determine the dimension of the key irreducible representation in every
specialization.
\end{abstract}
\vspace{.1in}
\noindent

In two dimensional statistical mechanics the Temperley-Lieb algebra has
exceptional structure when $Q=4 \cos^2(\pi/r)$ with $r$ rational. These
special cases are highly significant in several areas, including
representation theory,
conformal field theory and exactly solvable models
\cite{various,Jones,Cardy}.
In higher dimensions the analogue of the Temperley-Lieb algebra can be
 formulated \cite{pp8},
and the same question - what are the exceptional $Q$ values? - asked.
Here we give the answer for the limit of very high dimensions ( the ``mean
field'' case).

In \cite{pp8} we
introduced the partition algebra $P_n=P_n(Q)$. This is the algebra
represented by the single interaction transfer matrices of the
$n$ site, $Q$ state Potts
model
(in the dichromatic or Whitney polynomial realization \cite{Baxter})
 in arbitrarily high transverse dimensions (i.e. when every site in a transfer
matrix layer is a nearest neighbour of every other).

\noindent
Let us review this briefly:

For $M$ a finite set let $S_M$ be the set of partitions of $M$.
We may regard each partition $A \in S_{M}$
 as an equivalence relation on $M$
(and vice versa). For example, 
for $M= \{ 1,2,3 \}$
\[
A=((12)(3))=\{(1,1),(2,2),(3,3),(1,2),(2,1) \}
{}.
\]
If $M,N$ are two sets and $\mu,\nu$ equivalence relations on
$M$ and $N$ respectively, then define $\mu \circ \nu$ as the
equivalence relation on $M \cup N$ obtained by the transitive extension
of $\mu \cup \nu$ (e.g. $((12)(3)) \circ ((14)(5)) = ((124)(3)(5))$).

Now for given $n$ define sets
\[
[a] =\{a_1,a_2,..,a_n  \} 
\hspace{.6in}
[a;b]=\{a_1,a_2,..,a_n,b_1,b_2,..,b_n \}
\]
\[
[a;b;c]=\{a_1,a_2,..,a_n,b_1,b_2,..,b_n ,c_1,c_2,..,c_n \}
\]
and so on.    
We write $A[c;d]$ for the image of $A[a;b] \in S_{[a;b]}$ under the obvious
isomorphism
$J:S_{[a;b]} \rightarrow S_{[c;d]}$.
Note that there is a surjection
\[
R_c : S_{[a;c;b]}
          \rightarrow S_{[a;b]}
,
\]
 obtained by erasing all the $c_i$'s, and  a homomorphism
$f_c :  S_{[a;c;b]}
          \rightarrow \Z_{\ge 0}$ where $f_c(C[a;c;b])$ is
the number of equivalence classes of
$C[a;c;b]$ involving only $c_i$'s (for example,
$f_c(((a_1a_2b_1)(b_2c_1)(c_2)))=1$).

The partition algebra $P_n(Q)$ (over rational functions in $Q$)
may now be defined. It has basis $S_{[a;b]}$, and
multiplication of these basis elements is given as follows.
Consider the product of two partitions $A[a;b] * B[a;b]$.
Note that
$C[a;c;b]=A[a;c] \circ B[c;b] \in S_{[a;c;b]}$.
Then  $A[a;b] * B[a;b]
= Q^{f_c(C[a;c;b])} R_c(C[a;c;b])$.   

To make contact with the Potts model we note that the elements
\begin{equation}
A_{ii+1} = \sqrt{Q} \;\; ((a_1 b_1)(a_2 b_2)...(a_{i-1} b_{i-1})(a_i b_i
a_{i+1} b_{i+1})(a_{i+2} b_{i+2})...(a_n b_n))
\end{equation}
and
\begin{equation}
A_{i.} = \sqrt{Q^{-1}} \;\; ((a_1 b_1)(a_2 b_2)...(a_{i-1} b_{i-1})(a_i)(b_i)
(a_{i+1} b_{i+1})...(a_n b_n))
\end{equation}
obey the relations for generators of a Temperley-Lieb algebra (see \cite{pp8}
for more details).
Note also that
\begin{equation}
E_0 = \prod_{i=1}^n A_{i.}
\end{equation}
is, up to normalisation, a primitive central idempotent.

We found
in \cite{pp8}
a complete set of generically simple modules of
$P_n(Q)$, each equipped with an inner product.
These modules are indexed by a non-negative integer $i \in \{0,1,2,..,n\}$
and a partition $\lambda \vdash i$ (or more succinctly just by $\lambda$).
Let us write ${\cal S}_{\lambda}$ for such a module, and $M_n(\lambda)$
for the Gram matrix of the inner product.

For $i=0$ there is just the trivial partition $\lambda=0$. It is easy to
check that the corresponding left module $P_nE_0$ induces the
irreducible representation of the partition algebra
responsible for the transfer matrix sector containing the partition function
(given by the largest eigenvalue in the real temperature region).
Thus it is the single most important module from the physical point of view.
To see this recall \cite{pp8,Baxter,pp0}
that the transfer matrix may be written in the form
\[
T \propto \left( \prod_{i=1}^n (x+A_{i.}) \right)
      \left( \prod_{i\ne j} (1+xA_{ij}) \right)
\]
where $x$ is real for physical temperatures
and the matrix representation of the $A$ operators depends on
the precise details of the model.
In any case, for very small $x$ this $T$ is dominated
by a term in $E_0$. But $E_0$ vanishes in any irreducible representation other
than
${\cal S}_0$, since $E_0$ is a primitive idempotent for that representation.
Thus
for small $x$ the largest eigenvalue of $T$
must arise in this sector. On the other hand
$T$ is a positive matrix (of Boltzmann weights)
for all real $x$, so the largest eigenvalue is
never degenerate, by the Perron-Frobenius theorem. Thus the  eigenvalue largest
at small $x$ is the same one as that largest  at any other real $x$ value.

In this paper we focus on this module and determine its exceptional
structure (that is, at the $Q$ values where it ceases to be irreducible).
This is of particular interest in colouring problems \cite{pp8,Baxter,KS}, and
in the search
for solvable statistical mechanical models above two dimensions (c.f.
\cite{var2}).
Using some category theory we can then deduce the exceptional values of $Q$ for
the whole algebra.

Let $x = \sqrt{Q}$, $\lambda'$ be the conjugate partition to
$\lambda$, and (for $Q>0$)
\begin{equation}
\label{LL}
L_Q =L_Q(n) = \; card \{
A \in S_{[a;b]} \; | \; \mbox{ no. of parts of $A$ is} >Q \}
{}.
\end{equation}
Our key result is
\begin{equation}
\label{KK}
det \; M_n(0) = x^{L_0} \prod_{R=1}^{n-1} \left(
  (x + \sqrt{R})(x - \sqrt{R}) \right)^{L_R}
\end{equation}
where $L_0$ is a known positive integer.

\noindent
{\em Proof:}

For given $n$,
in this $ P_n E_0$        
space the basis states
are in one-to-one correspondence with the partitions of $[a]$. 
For example with $n=3$, and writing just $i$ for $a_i$ and
$i'$ for $b_i$, we have a basis
\[
B_p = \{
((123) \;     (1')(2')(3')), \hspace{3in}
\] \[
((12)(3) \;   (1')(2')(3')), \;
((1)(23) \;   (1')(2')(3')), \;
((13)(2) \;   (1')(2')(3')), \;
((1)(2)(3) \; (1')(2')(3')) \;\; \}
{}.
\]
To avoid overcounting we have adopted the convention of writing partitions
with the lowest possible (unprimed)
numbers first, as above (e.g. always start with 1).
This is a total order.

Another useful basis (differing from the
partition basis only by factors of $x$)
is in terms of words in the generators $A_{i.}$
and $A_{ij}$:
\[
B_w =\{
E_0,        \;
A_{12} E_0, \;
A_{23} E_0, \;
A_{13} E_0, \;
A_{12} A_{13} E_0  \}
{}.
\]
For the sake of uniqueness we take the lowest possible indices in forming
these words (for example $A_{12}A_{23}E_0=A_{12}A_{13}E_0$ and we take the
latter). The correspondence between the two bases is then immediate (and
we will often speak of elements of $B_w$ as if they are the
corresponding partitions, but ignoring the primed elements which play no
role here).
For example
\[
x^{-3} ((157)(236)(4)...) \cong A_{15}A_{17} \left(  A_{23}A_{26} \right) E_0
\]
and generally
\[
x^{(-n+s+t+u+...)}((1p_1p_2..p_s)(q_0q_1..q_t)(r_0r_1..r_u)....) \cong
  \left( \prod_{k=1}^s A_{1p_k} \right)
 \left( \prod_{k=1}^t A_{q_0q_k} \right)
\left( \prod_{k=1}^u A_{r_0r_k} \right) .... E_0
\]
The partitions/words in
any $S_M$ may be partially ordered by the number of parts, and then
further sorted by their shapes (as partitions of $2n$ in the Young diagram
sense
\cite{Mac}, i.e. ignoring their the precise content of each part but
only noting its size). If $E_0$ (with $n$ parts and shape $(1^n)$, ignoring
primed
elements) is the first in the order then
\begin{equation}
\label{WP}
W_p =  \left( \prod_{i=2}^n A_{1i} \right)  E_0
\end{equation}
(one part, shape $(n)$) is the last.

If all the words $W_i$ in the word basis are written with their letters in
reverse order (denoted $W_i^T$)
then we get a basis for an (isomorphic) right module.
Let $B_w = \{ W_i | i=1,2,..,{\cal P}_n \}$ be the word basis,
then the Gram matrix $M=M_n(0)$ is given by
\begin{equation}
\label{WW}
W_i^T W_j = M_{ij} E_0
{}.
\end{equation}
Now $det(M)$ is polynomial in $x$, and symmetric.

If $\Lambda$ is a lower uni-triangular matrix of dimension ${\cal P}_n$
and $W_i$ is regarded as the $i^{th}$ component of a column vector $W$, then
another basis is $B_{\Lambda} = \{ (\Lambda W)_i | i=1,2,..,{\cal P}_n \}$.
In this basis the Gram matrix is $\Lambda M \Lambda^{\dagger}$, but
$det \; M$ is unchanged. In particular there will be an orthogonal basis such
that
\[
( \Lambda M \Lambda^{\dagger} )_{ij} =  \delta_{ij} \; F_i(Q)
\]
with $F_i(Q)$ a  non zero function, so
\begin{equation}
\label{KK2}
det \; M = \prod_i F_i(Q)
{}.
\end{equation}
If $P_n E_0$ develops a proper invariant subspace of dimension $d$ (say)
in some specialisation $Q=Q_c$, then $d$ of these functions vanish at
$Q=Q_c$.

Note that $M_{ii}=x^n$. To see this note that in particular
\begin{equation}
\label{use}
E_0E_0=x^n E_0
\end{equation}
and that $[A_{ij},A_{kl}]=0$. Thus suposing the proposition true
for $W_h$
and working by induction on
the length of word then the next word may be written
$W_h^T A_{jk} \left( A_{jk} W_h \right) = x W_h^T A_{jk} W_h$.
On the right hand side we have a factor $xA_{k.} A_{jk} A_{k.} = xA_{k.}$
compared with $A_{k.} A_{k.}=x A_{k.}$ in
$W_h^T W_h =x^n E_0$,
so altogether the right hand side is unchanged.

Furthermore writing $M_{ij}=x^{m_{ij}}$ then $m_{ij}<n$ unless $i=j$.
This is because $i \neq j$ implies that
at least some of the $A_{kl}$-type factors are
different in $W_i^T$ and $W_j$, and hence only occur once in
$W_i^T W_j$. Now suppose $A_{kl}$ appears in $W_j$ only. Either it
would be redundant in $W_i^T$, e.g. if $A_{km}A_{ml}$ was there,
in which case it may be replaced by one of these in the product $W_i^T W_j$
(by the definition of the algebra product),
or it would not. In either case by inserting some extra factors
into  $W_i^T W_j$ we
can obtain $W_h^T W_h$
(in which every factor appears twice)
for some longer word $W_h$. But each such extra factor
contributes a factor $x$ to $M_{hh}=x^n$, so the exponent of $M_{ij}$ is
smaller by the number of these factors.

Consequently $det(M)$ is of degree $n . {\cal P}_n$.

Now note that the exponents $L_R$ in equation~\ref{KK} are lower bounds. This
follows
since a $Q$ state system ($Q$ a positive integer)
cannot support more than $Q$ distinct parts on the
complete graph (in the $Q$ colouring interpretation \cite{pp8} nodes are
connected - in
the same part -
if they are adjacent and coloured the same, but on the
complete graph all nodes are adjacent, so at
most $Q$ parts are possible, one for each different colour).
Thus for $Q$ a positive integer the basis states in $B_w$ corresponding to
partitions of more than $Q$ parts cannot be linearly independent of
the rest. An explicit proof of this also exists - for the
sake of brevity, let us simply take an example:
the case $n=2$. Here we have $B_w
= \{ E_0, A_{12} E_0 \}$, and $(1-A_{12})E_0$ spans a one dimensional
invariant subspace when $Q=1$. Quotienting by the invariant subspace
introduces a linear dependence between the two states (see also \cite{KS}).

In other words $L_{Q_c}$ of the factors in equation~\ref{KK2} must vanish at
$Q=Q_c$ (a positive integer).
Since these factors are rational in $x$ they vanish like
$(x^2-Q_c)^{\alpha}$ where $\alpha$ is a positive integer.
Hence the total degree in $x$ of $det \; M$ is
\begin{equation}
\label{LX}
L_0 +2.\sum_{R=1}^{(n-1)} L_R        
  +X = n.{\cal P}_n
\end{equation}
where 
$X \ge 0$ is the contribution of other factors not given in equation~\ref{KK}.
To show $X=0$ we compute $L_0$, which is just the lowest
overall power of $x$ of any term in the determinant.

Since each $A_{jk}$ factor in $W_j$ reduces the number of parts by 1 the
maximum number of such factors is $n-1$
(specifically in $W_p$).
To minimize
$m_{ij}$ we want the minimum number of duplicated factors in
$W^T_i W_j$, but the maximum number of single factors.
For example
\[
E_0 W_p = x E_0
\]
gives the (equal) smallest possible exponent overall.
 From equation~\ref{WP} this has $n-1$ $\; A_{jk}$ factors between
two $E_0$s.
Any fewer
$A_{jk}$ factors and some identical pair of $A_{i.}$ factors, one from the
$E_0$ on
each side, necessarily meet.
Any more and some identical pair of $A_{jk}$ factors can be made to meet.
Either situation gives extra factors of $x$ on the right hand side of
equation~\ref{WW}.
Now each term in the determinant involves juxtaposing each of the
${\cal P}_n \;$
$W_i^T$s with some $W_j$, so in each term all the $W_i^T$s and
$W_j$s contribute once each. The minimum number of $x$ factors overall would
arise if each $W_i^T W_j$ combination could be arranged to have
$n-1$ appropriately distinct $A_{jk}$ factors (i.e so that $W_i^T W_j = E_0
W_p$).
Altogether that would require exactly $(n-1).{\cal P}_n$ factors,
and we would have a lower bound for $L_0$ of ${\cal P}_n$ (just one factor of
$x$ from each pair).
Furthermore, it follows that if $(n-1).{\cal P}_n$
 is {\em not} the total number present then
the discrepancy is a lower bound on $L_0 - {\cal P}_n$
(the bound is realized if all the individual discrepancies are
coherent, and otherwise all the factors can be arranged in the
optimum way - checking these two things out explicitly would be tough, but in
fact
the bound is already saturated, as we will see).

In fact the total number of $A_{jk}$ factors in $B_w$ (or $B_w^T$) is
\[
Z = \sum_{\lambda \vdash n} D_{\lambda} \left( n- \lambda_1' \right)
\]
where $D_{\lambda}$ is the number of partitions in $S_{[a]}$ of shape $\lambda$
\cite{Liu,pp8}, i.e. $\sum_{\lambda} D_{\lambda} = {\cal P}_n$. This is
because $\lambda_1'$ is the number of parts for any partition of shape
$\lambda$;
a partition of $n$ parts has no $A_{jk}$ factors (it is just $E_0$);
and each
$A_{jk}$ factor reduces the number of parts in a partition by one.
Altogether then we have $2 Z$ factors.
This gives an excess $2Z - (n-1)).{\cal P}_n$, so
$L_0 \ge 2Z-(n-2)).{\cal P}_n$.

On the other hand
\[
 \sum_{R=1}^{n-1} L_R
  =
 \sum_{\lambda \vdash n } D_{\lambda} \left( \lambda_1' -1 \right)
\]
since a partition with $\lambda_1'$ parts counts in $L_R$ for
all $R=1,2,..., \lambda_1' -1$. From equation~\ref{LX}
this gives
  $L_0 \le n.{\cal P}_n  - 2 .
\sum_{\lambda \vdash n } D_{\lambda} \left( \lambda_1' -1 \right)$.
The reader will now readily confirm that the bounds meet,
and so $X=0$.

QED.

\vspace{.1in}

Consequently the module ${\cal S}_0$ is simple unless $Q$
is a natural number less than $n$
(in which case the irreducible dimensions are ${\cal P}_n - L_Q$).

Also, if
all ${\cal S}_{\lambda}$ for $| \lambda | < i$
are simple at level $n$ for some $i$ then
all ${\cal S}_{\lambda}$ for $| \lambda | < i+1$
are simple at level $n-1$
by Frobenius reciprocity \cite{mathfool}
(and by using the induction/restriction rules and category
properties of $P_n(Q)$ given in \cite{pp8}).
In particular if
${\cal S}_0$ is simple at level $2n$, say, (case $i=1$) then
${\cal S}_{\lambda}$ is simple for all $\lambda$ at level $n$, and so
$P_n(Q)$ is semi-simple.
Since ${\cal S}_0$ is simple for all $n$ for all $Q \not \in \N$
then $P_n(Q)$ is semi-simple for all $n$ for $Q \not \in \N$.

\smallskip

Our result rises some questions. From the physics point
of view, one may wonder what kind of symmetries do appear in the mean
field Potts model
for $Q$ an integer, and whether the degeneracy in the corresponding
Temperley Lieb algebra
indicates the existence of new models of ``restricted type''. Recall that in
the case of two dimensional statistical mechanics the exceptional values
$Q=4\mbox{cos}^2(\pi/r)$ were associated with rational conformal field
theories
, and the existence of restricted solid on solid models \cite{various}.
In the mean field case, the standard Potts model has
 first order phase
transition for $Q>2$. Hence the presence of  symmetries (for
 $Q$ an integer) should manifest itself rather differently
than in two dimensions where there are second order phase transitions and a
 continuum limit with degenerate Virasoro algebra representations. From a
more mathematical point of view, one can wonder whether
the striking relation between degeneracy of the Temperley Lieb algebra
and zeroes of chromatic polynomials, observed so far in the planar case
\cite{KS} and now in the mean field case also, generalizes to finite
dimensions.


\begin{thebibliography}{99}
%
\bibitem{various} Andrews G E,  Baxter R J and Forrester P J, J.Stat.Phys. 35
(1984) 193;
Huse D.A., Phys.Rev. B30 (1984) 3908;
Belavin A A, Polyakov and  Zamolodchikov A B , Nucl.Phys.
B241 (1984) 333;
Date E,  Jimbo M ,  Miwa T  and  Okado M, Nucl.Phys. B290 [FS20] (1987) 231;
Deguchi T, Wadati M  and  Akutsu Y, J.Phys. Soc. Japan 57 (1988) 2921;
Pasquier V. and
Saleur H, Nucl.Phys. B330 (1990) 523, and references therein.
\bibitem{Jones} Jones V F R, Invent.Math. 72 (1983) 1.
; Dipper R. and G. James, Proc.London Math.Soc. 52(1986) 20 and refs. therein.
%
%
\bibitem{Cardy} Cardy J, in {\em Phase transitions and critical phenomena}
vol.11,
eds. C.Domb and J.Lebowitz (Academic Press).
%
\bibitem{pp8} Martin P P, Non-planar statistical
mechanics - The Partition Algebra construction, Yale preprint YCTP-P34-92,
submitted to Comm.Math.Phys.; $\;$
                 Martin P P and H Saleur, On an Algebraic Approach to
Statistical Mechanics
in Higher Dimensions, Yale preprint YCTP-P33-92, submitted to Comm. Math.
Phys..
\bibitem{Baxter} Baxter R J, {\em Exactly Solved Models in Statistical
Mechanics},
(Academic Press, New York, 1982).
\bibitem{KS}Saleur H , Comm. Math. Phys. 132 (1990) 657; Kauffman L,
Saleur H, Comm. Math. to appear
Phys.
\bibitem{pp0} Martin P P, 
                          Publ.RIMS Kyoto Univ. 26 (1991) 485.
%
\bibitem{var2} Zamalodchikov A B, Comm. Math. Phys. 79 (1981) 489;
 Baxter R J , in {\em Integrable systems in statistical mechanics}, ed.
M.Rasetti et al
(World Scientific, Singapore, 1985);
 Baxter R J  and  Bazhanov V V , J. Stat. Phys. to appear. 
%
%
\bibitem{Mac} MacDonald I G, {\em Symmetric Functions and Hall Polynomials},
(OUP, Oxford, 1979).
\bibitem{Liu} Liu C L, Introduction to Combinatorial Mathematics,
McGraw-Hill,NY, 1968.
\bibitem{mathfool} see, for example, Cohn P., {\em Algebra} vol.2, (John Wiley,
London, 1989).
\end{thebibliography}
\end{document}